\documentclass[dvips,11pt]{report}
\usepackage{lscape,graphicx}
\usepackage{color}
\usepackage{amssymb}
\usepackage{euler}
\def\bmath#1{\mbox{\boldmath$#1$}}
\newcommand{\huts}{Hutsem\'{e}kers }
\begin{document}




\begin{center}
{\large\bf 
Interpretation of the Global Anisotropy in the 
Radio Polarizations of Cosmologically Distant Sources}

\bigskip
{\large\bf Pankaj Jain and S. Sarala}

\bigskip
{Physics Department, IIT, Kanpur, India 208016}
\end{center}

\bigskip
{\bf Abstract:} We present a detailed statistical study of the observed 
anisotropy in radio polarizations from distant extragalactic objects.  
This anisotropy was earlier
found by Birch (1982) and reconfirmed by 
Jain \& Ralston (1999) in a larger data set. A very strong signal
was seen after imposing the cut $|RM-\overline{RM}|>6$ rad/m$^2$, 
where $RM$ is the rotation measure and $\overline{RM}$ its mean value. 
In this paper we show that there are several indications that
this anisotropy cannot be attributed to bias in the data.
 We also find that a generalized statistic 
shows a very strong signal in the entire data without imposing 
the RM dependent cut. Finally we argue that an anisotropic
background pseudoscalar field can explain the observations.


\newpage
\noindent
{\Large\bf 1. Introduction}

\medskip
Polarizations of radio waves from extragalactic 
sources undergo Faraday Rotation
upon propagation through galactic magnetic fields.
This effect provides very useful information about 
astrophysical magnetic fields 
(Zeldovich, Ruzmaikin \& Sokoloff 1983, Vall\'{e}e 1997).
The amount of rotation is proportional to the
magnetic field component parallel to the direction
of propagation of the wave and to the square of the
wavelength $\lambda$. The observed orientation $\theta (\lambda)$ of
the linearly polarized component of the electromagnetic
wave  can therefore be written as,
\begin{eqnarray}
\theta (\lambda^2) = \chi + \left( RM
\right)\lambda^2
\label{PAfit}
\end{eqnarray}
where the slope, called Faraday Rotation Measure ($RM$), depends 
linearly
on the line integral of the parallel component of the magnetic field 
along the direction of propagation of the wave and $\chi$ is
the intercept, also called the intrinsic position angle of 
polarization, $IPA$. 

The observed polarization angle $\chi$, after the 
effect of Faraday rotation is taken out of the data, 
is observed to be dominantly aligned perpendicular to the
orientation axis of the galaxy. Let $\psi$ denote the orientation
angle of the galaxy. Then
the offset angle $\beta= \chi-\psi$ is found to be approximately 
equal to $\pi/2$ for most of the sources, i.e. the distribution of
$\beta$ over a large sample of sources is found to peak at $\beta\approx 
\pi/2$. Besides this dominant trend, a smaller peak is also found at
$\beta \approx 0$. This suggests the existence of 
two populations, some with polarization position angles
parallel and others perpendicular to the galaxy axis.

In $1982$, Birch empirically observed an angular
anisotropy in the offset angle $\beta$, using a data
set of $137$ points. 
Birch's statistics were questioned
by Phinney \& Webster (1983) and it was pointed out that the
significance of Birch's result can be significantly reduced if the
experimental errors in $\beta$ are taken into consideration.
Phinney \& Webster (1983) also suggested that the signal observed
by Birch might result from the presence of bias in data. 
Kendall \& Young (1984) further investigated Birch's claim of 
cosmic anisotropy with more sophisticated statistics and using an 
updated version of
Birch's data. They found 
that the statistics were not consistent with isotropy at $99.9$\%
confidence level. Later Bietenholz \& Kronberg (1984) repeated the
calculations using single-number correlation test statistic,
originally proposed by Jupp \& Mardia (1980).
This statistic also showed strong evidence of anisotropy 
in Birch's data with a confidence
level of $99.98$\%. They went on to create an independent set of 277 points
which, however, showed 
no signal of anisotropy. This lead to a dismissal of 
Birch's results but left  
unresolved the puzzling fact that his data had
contained a signal of anisotropy at such a high level of statistical
significance.

The possible existence of anisotropy in radio polarizations was 
reanalysed by
Jain \& Ralston (1999). The authors
collected an independent set of 
$361$ points from all the available catalogues. This set included
$\beta$ values for 
29 sources which were contained only in the 
Birch's compilation. Jain \& Ralston (1999) also considered the data set of
332 sources obtained after deleting these 29 objects. The authors found
that both the data sets showed a statistically significant
signal of anisotropy. The observed signal can be expressed as follows,
\begin{equation}
2\beta - \pi \approx \bmath{\Lambda}\cdot \bmath{R}
\label{eq:anisotropy}
\end{equation}
where \bmath{R} is a unit vector in the direction of the source. The
vector \bmath{\Lambda} represents the 
three parameters of this fit and points in the prefered direction of this
dipole anisotropy. 
The unit vector \bmath{\hat \Lambda}  
controls the distribution of $(2\beta-\pi)$ 
on the dome of the sky 
such that it is predominantly positive in the direction of the axis 
while it is predominantly negative in the opposite 
direction.

Jain \& Ralston (1999) tested the isotropy 
as null hypothesis using Maximum Likelihood analysis by taking into account 
the transformation property of $\beta$. 
A typical null distribution for angular
variables is the von Mises (vM) distribution 
\begin{equation}
vM(\theta) = const \times e^{k \cos(\theta)}
\label{eq:vm}
\end{equation}
where $k$ is a constant and $\theta$ is the angular parameter which
is taken here to be $2 \beta$. The factor 2 arises since $\beta$ refers
to the polarization angle and $\beta+\pi$ is identified with $\beta$ 
(Ralston \& Jain 1999). 
The joint distribution  
of the angle $\beta$ and source positions was taken to be of the
form $$f(2\beta)C(2\beta,{\bf R})
h({\bf R})\ ,$$ where $f(2\beta)$ is the marginal distribution of
$\beta$ which is assumed to be the von Mises,
$h({\bf R})$ is the angular distribution of the sources
and the correlation 
ansatz is taken to be of the form,
\begin{equation}
C(2\beta,{\bf R}) = \exp\left[\bmath{\lambda}
\cdot {\bf R} \sin{2\beta}\right]
\label{joint}
\end{equation}
Here \bmath{\lambda} represent the three parameters of the correlation
ansatz.
An alternate distribution, obtained by replacing 
\begin{equation}
2\beta\rightarrow
\Omega(\beta) = 
2\beta + \nu sin 2\beta
\label{transformation}
\end{equation}
was also considered by Jain \& Ralston (1999) since
it was found to provide a better null fit to the data.
The fit selected a negative value of the parameter $\nu\approx -1$ which
leads to a more sharply peaked distribution in comparison to the von
Mises. Such a sharply peaked function is also justified 
by the distribution of polarizations obtained under the
assumption that the electric field components of the electromagnetic wave
are normally distributed (Sarala \& Jain 2001, 2002). 

\begin{figure}[b!]
\includegraphics[scale=0.85]{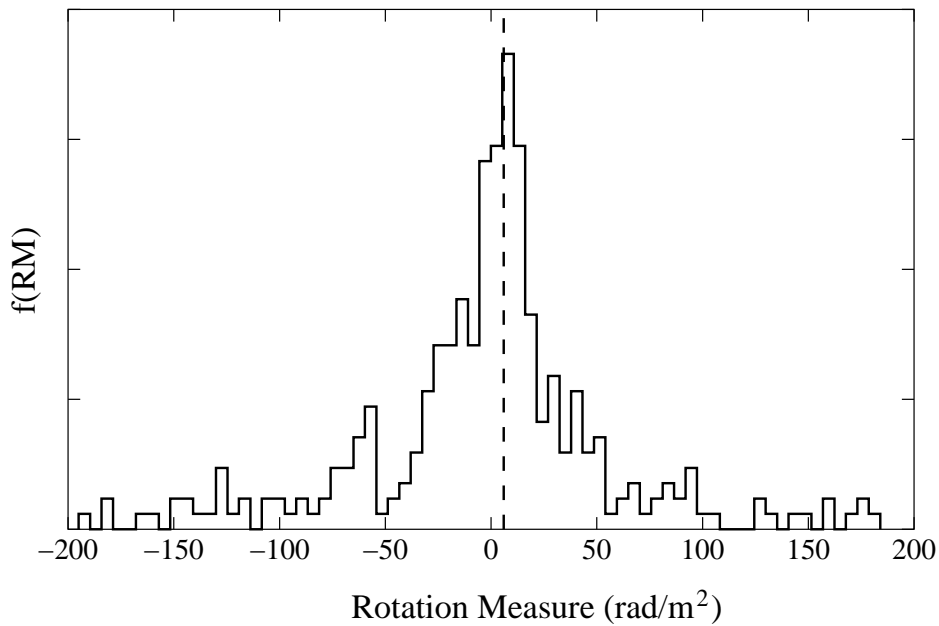}
\caption[The distribution of $RM$ for $332$ points,
excluding Birch data]
{The distribution of Rotation Measure for the data with $332$
points
excluding some of the sources listed in Birch (1982) 
for which information on $RM$ was not
available. The $RM$ distribution is
not exactly Gaussian. Its mean as well as the peak position is
shifted from $RM=0$ towards positive values, as discussed in
text. The vertical {dashed line} shows the position
of the mean, $\overline{RM} = 6$.}
\label{hist_rm}
\end{figure}

Likelihood analysis showed a strong signal of anisotropy using the
transformation \ref{transformation} 
in the
joint distribution \ref{joint}. The strongest signal was obtained
after making the cut on the rotation measures so as to eliminate
the data with RM in the range $|RM-\overline {RM}|\le 6$, where
$\overline {RM}= 6\ rad/m^2$ is the mean value of the rotation measure
over the data sample. 
In fig. \ref{hist_rm} we show the histogram of the rotation measures 
for the 332 sources. It is found that the peak position as well as
the mean value of RM is shifted from zero and
lies approximately at $6\ rad/m^2$. 
This is somewhat unexpected since
for a large unbiased sample we would expect the peak value of
$RM$ to be zero. 
Even for a larger data sample compiled in 
Broten, MacLeod \& Vallee  (1988) 
we find similar distribution with peak position
as well as the mean and median of the rotation measures shifted
from zero towards positive values. 
The cut on RM, therefore, removes points lying close to the central peak in the 
distribution. The statistical significance of the signal after this
cut was found to be approximately 0.06\%, i.e. the probability that
the correlation seen in data might arise in a random sample is
given by $P=0.06$\%. The unit 
vector \bmath{\hat\lambda} after making the cut  $|RM-\overline {RM}|>6$
was found to be (Jain \& Ralston 1999)
\begin{equation}
\bmath{\hat \lambda} = [(0\ h, 9\ m)\pm (1\ h,0\ m),-1^o\pm 15^o]\ 
\label{axis}
\end{equation} 
in the B1950 coordinate system.

Similar results were obtained by using the Jupp and Mardia correlation
coefficient. In this case the authors (Jain \& Ralston 1999) 
found the correlation between
$\sin(2\beta - sin 2\beta)$ and the angular position vector of
the source. The parameter $\nu$ in Eq. \ref{transformation}
was set equal to $-1$  
since the likelihood analysis generally prefered a
value between $-0.5$ and $-1$. 
The statistical significance in this case, after 
making the cut on rotation measures, was found to be $P=0.04$\%. 

In the present paper we further illustrate the nature of this correlation.
The basic cause of this correlation is so far unknown. There have
been several proposals such as observational bias, bias in the 
extraction of RM and $\beta$ (Phinney \& Webster 1983)
existence of a light pseudoscalar
(Jain, Panda \& Sarala 2002), cosmic rotation (Obukhov 2000,
Obukhov, Korotky \& Hehl 1997,
Mansouri \& Nozari 1997),
inhomogeneous and anisotropic universe (Moffat 1997),
weak lensing by large-scale density inhomogeneities (Surpi \& Harari 1999),
background torsion field (Dobado \& Maroto 1997)
etc. The fact 
that the correlation is improved significantly after making a cut
based on rotation measure may provide a hint about its origin. One
obvious possibility is that there may exist bias in the extraction
of RM and $\beta$ which leads to a correlation between these two parameters. 
Then, since
the $RM$ or the magnetic field is not isotropically distributed,
it can induce an anisotropy in $\beta$.  
Indeed such a possibility 
was suggested (Phinney \& Webster 1983), soon after the appearance of the Birch's paper.

\bigskip
\noindent
{\Large\bf 
2. Statistical Analysis of the Angular Correlation in Polarizations}

\medskip
We next perform several statistical tests in order to determine whether
the observed correlation can be explained in terms of bias in the values
of $\beta$ which may result from   
biased extraction of rotation measures. As discussed above if the extraction
of RM and $\beta$ is biased then since the RM is not isotropically distributed
on the celestial sphere it can lead to an anisotropic distribution of
$\beta$. In order for this mechanism to work 
it is necessary that $\beta$ is strongly correlated with $RM$. 
We next examine if such a correlation exists in the data by
computing the
Jupp \& Mardia (1980) (JM) correlation between $RM$ and
$\beta$ for the data set containing $332$ sources. The correlation is
found 
to be very small with $n\rho^2 = 1.742$. We
tried several cuts to see whether the correlation is enhanced and 
found a small correlation with $n\rho^2 = 4.341$ for 
$|RM| < 100$ rad/$m^2$ with the number of sources $n=278$. 
A more stringent cut on $RM$ does
not improve the correlation. We summarize JM
correlation with different cuts in Table \ref{JM_RM_beta}. 
We see from this table that $RM$ is not correlated with
$\beta$. This implies that a large range of explanations of the
anisotropy which invoke the possibility of biased estimation
of $RM$ and $IPA$, including the mechanism proposed by Phinney
\& Webster (1983), are disfavored. 

\begin{figure}[h!]
\includegraphics[scale=0.85]{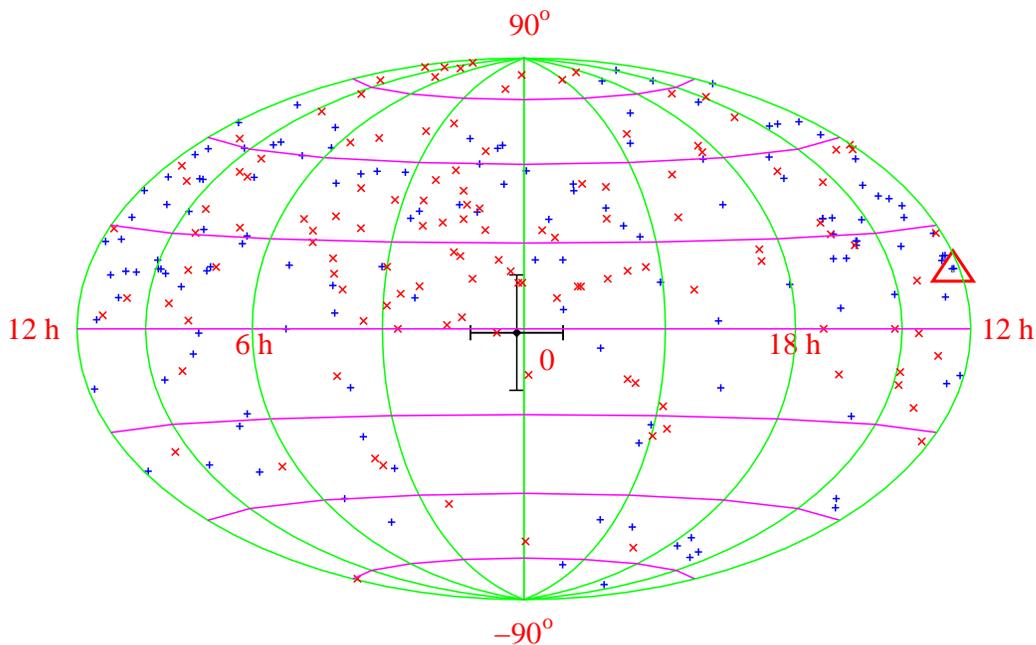}
\caption[Aithoff-Hammer equal-area plot of $\sin 2 \beta$ on 
the dome of the sky ]{Aithoff-Hammer equal-area plot of 
$\sin 2 \beta$, where $\beta$ is the
difference in the polarization position angle and the
orientation angle of the source.  
We represent $\sin 2 \beta <0.0$ by blue pluses while 
$\sin 2 \beta \ge 0.0$
is represented by red crosses. The dipole anisotropy axis, 
$[(0\ h, 9\ m)\pm (1\ h,0\ m),-1^o\pm 15^o]$, obtained 
by Jain \& Ralston (1999)
is shown along with its error bars. The position of the 
Virgo supercluster
is shown with a red triangle. The cut $|RM-6|> 6$ has
been used in this plot.}
\label{aniso}
\end{figure}

The above lack of correlation between $RM$ and $\beta$, however,
does not rule out the possibility that the value of $\beta$ is 
randomized due to biased extraction of $RM$. 
This might happen
preferentially in the regions of large magnetic field i.e. where 
$RM$ is large. As we have discussed
earlier, the distribution of $\beta$ peaks at $\pi/2$. 
If in a certain patch the values of $\beta$ are randomized,
then the mean value of $\beta$ in this patch will deviate 
significantly from $\pi/2$ and could
in principle give rise to a large value of the statistic
indicating correlation. 

\begin{figure}[h!]
\centering
\includegraphics[scale=1.0]{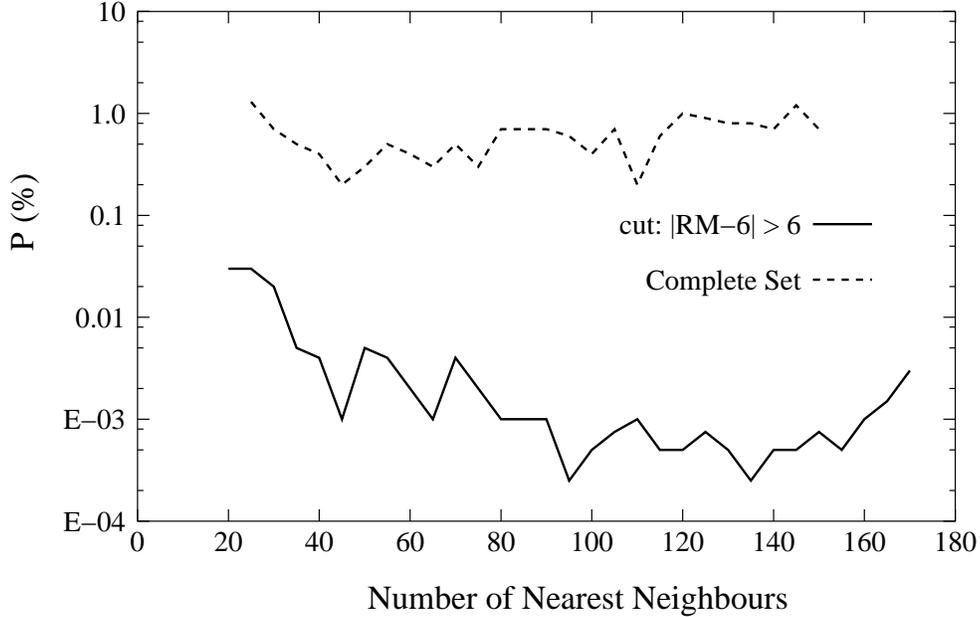}
\caption[Significance level using $X$ statistic]
{The $P$-value(\%) using the $X$ statistic is plotted
as the function of the number of nearest neighbours. 
The result is shown
for both the complete set($332$ points) and with the cut in the
$RM$ ($265$ points).}
\label{fig:sign}
\end{figure}

\begin{figure}[h!]
\centering
\includegraphics[scale=0.95]{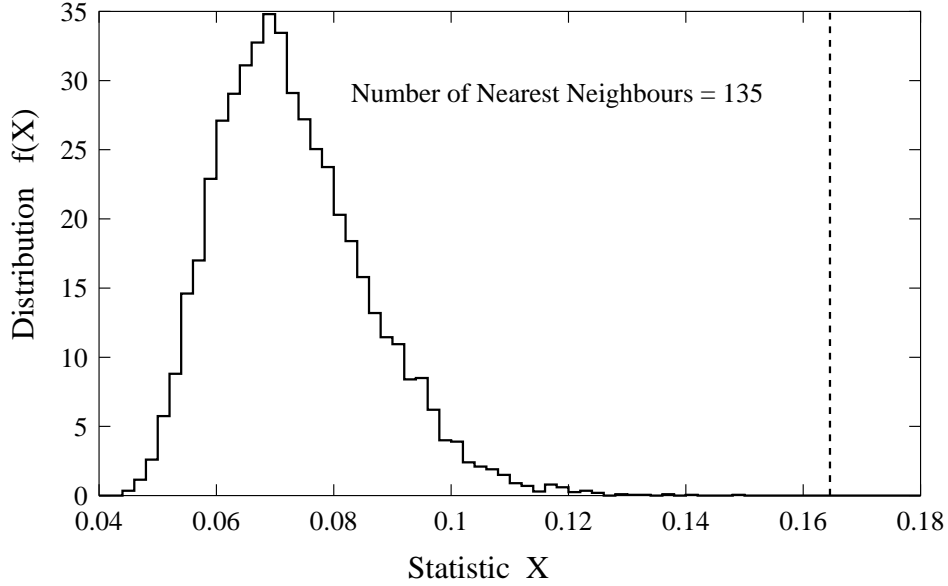}
\caption[Histogram of $X$ statistic]
{The histogram of $X$ statistic using 10000 random
samples obtained by shuffling the $\beta$ values of the data with
the cut $|RM-6|>6$. The
number of nearest neighbours is taken to be 135 at which the
$P$-value is found to be the smallest. The statistic of the data in
this case is equal to 0.165, shown by the dotted line, and the
corresponding $P=2.5\times 10^{-4}$ \%.  }
\label{hist:sign}
\end{figure}

We next investigate whether the correlation arises due such a randomization. 
In Figure \ref{aniso} we plot the sign of 
$\sin 2 \beta$ over the dome of the sky
in the equatorial coordinate system after making the cut
$|RM-6|>6$. 
Here the sources with $\sin 2 \beta <0.0$ and 
$\sin 2 \beta \ge 0.0$
are represented by plusses and crosses respectively. 
From this figure we see
that the positive values of $\sin 2 \beta$ dominate near the dipole
anisotropy axis given in Equation \ref{axis}. Figure \ref{aniso} clearly 
indicates that there indeed exists an anisotropy in the 
polarizations of radio galaxies and
there exist large patches where the value 
of $\sin{2\beta}$ is either positive or negative. Hence we find
no evidence that the correlation seen in data arises due to
randomization of $\beta$ in some regions. 
We may quantize this clustering of negative and positive values
of $\sin2\beta$ by defining a new statistic. We consider
a $N_n$ number of nearest neighbours of the source $i$. 
For these $N_n$ sources we set $x_i=1$ if $\sin2\beta\ge 0$
and $x_i=-1$ if $\sin2\beta<0$. We then compute
$X_i = |\sum_i x_i|$ and the statistic $X$ is obtained by summing
$X_i$ over the entire data sample. The $P$-value in this case
is computed by comparing with a large number of random samples
obtained by shuffling the $\beta$ values of the data. The
resulting $P$-values or the significance level
for the complete data sample and for the set obtained after 
making the cut $|RM-6|>6$ are shown in Figure \ref{fig:sign}. 
We find that $P\approx 0.2 - 0.5$\% for the complete sample and 
$P\approx 2.5\times 10^{-4} - 6.0\times 10^{-4} $\% after the cut.
In Figure \ref{hist:sign} we show the histogram of the statistic
$X$ generated using 10000 random data samples. This figure also
clearly shows that the statistic of the data sample lies far
above the peak value of the histogram.
The very small $P$-values obtained after the cut is a clear
indication that $\beta$ values in the different regions of the
sky are indeed correlated with one another. 
This is infact
another confirmation of the large correlation found by Jain
\& Ralston (1999). 
Hence we dismiss the above explanation in terms of bias in the 
extraction of $IPA$. 

An improved statistical procedure for extraction of rotation measures
and IPA was proposed by Sarala \& Jain (2001, 2002). Using the
revised $\beta$ values obtained by this procedure, the Jupp-Mardia
test for anisotropy gives $P=0.1$ \%. Hence these revised
$\beta$ values continue to support the presence of
anisotropy in the data.

\begin{table}[h!]
\begin{center}
\begin{tabular}{|c|c|c|}
\hline \hline 
{\bf Cuts} &{\bf n }& \bmath{n \rho^2}\\
\hline \hline 
Full data &332 & 1.742\\
$|RM-6|>6 $ &265 & 1.780\\
$|RM|<100$ &278 & 4.341\\
$|RM|<70$ &254 & 1.101\\
$|RM|<100$ and $|RM-6|>6 $ &211 & 4.320\\
\hline \hline 
\end{tabular}
\end{center}
\caption{JM correlation between $RM$ and $\beta$ with
different cuts in the $RM$.}
\label{JM_RM_beta}
\end{table}

In order to further examine the effect of rotation measures
on the observed anisotropy we examine
how the correlation changes if we eliminate the sources which
lie close to the galactic plane. The dominant contribution to
$RM$ is obtained from the milky way and hence largest values of
$RM$ are observed for these sources. We therefore impose a cut
on the data to eliminate sources with the galactic latitude 
$|b|\le 30^o$. This leaves a total of 214 sources in the data
set. We repeat the JM correlation discussed earlier
using Map 3.
We find that the statistic $n \rho^2 =9.43$ after the galactic cut and
this implies that the anisotropy in $\beta$ persists with about 98\%
confidence level. As expected, the anisotropy is enhanced by making a cut in
the $RM$ such that only the sources which statisfy 
$|RM-6|>6.0$ are included. The results are
tabulated in Table \ref{gcut}. We see from the table that the
statistic does not change much in comparison with the data without 
the galactic cut. Hence we find that the regions of large $RM$ do
not necessarily imply large correlation of $\beta$.
\begin{table}[h!]
\begin{center}
\begin{tabular}{|c|c|c|c|}
\hline \hline 
& n & \( n \rho^2 \) & P (\%)\\ \hline \hline 
Full Set & 214& 9.43 & 2\\ \hline 
\textbf{\( |(RM-6|>6.0 \)}& 155& \textbf{22.33} & 
0.04$^*$\\
\hline \hline 
\end{tabular}
\end{center}
\caption[JM correlation after galactic cut]{
The $n \rho^2$ values after eliminating sources
lying in the galactic plane. Results
are given for the full set as well as with a cut
in the Rotation Measure.
$^*$The $P$ value for the cut $|(RM-6)|>6.0$
is obtained by using $\chi^2_5$.}
\label{gcut}
\end{table}

The above discussion clearly shows that the correlation seen
in the data is not a direct consequence of the correlation of the 
RM with the angular coordinates of the source. However the cut
in RM does result in a large increase in the signal of anisotropy
and hence in some sense it must depend on RM. One possibility
is that the sources eliminated by the cut show a random behaviour
and are uncorrelated. This turns out to be not true. Instead we find that
the sources eliminated by the cut show an opposite angular dependence
in comparison to the remaining sources. We illustrate this by making
the following ansatz for the joint distribution
\begin{equation}
F(\beta,{\bf R})h({\bf R}) = \exp\left[k\cos{\Omega} + 
\bmath{\lambda}\cdot {\bf R}(1+g(RM)) \sin{\Omega}\right] h({\bf R})
\label{joint_rm}
\end{equation}
where as defined earlier $\Omega=2\beta+\nu \sin 2\beta$ and 
$g(x)$ is taken to be a gaussian, 
\begin{equation}
g(x) = c_1\pi\exp\left[-(x-c_2)^2/c_3^2\right]\ .
\label{eq:gaussian}
\end{equation}
Here we have introduced three new parameters $c_1,c_2$ and $c_3$.
The difference of maximum likelihood between this correlated ansatz
and the null ansatz turns out to be 17.6 for the entire data set of 
332 sources. Here the null ansatz
is the two parameter distribution obtained by using the transformation 
\ref{transformation} in the von Mises distribution. This ``sharply
peaked" distribution can be written as
\begin{equation}
S(\beta) = const \times e^{k \cos(2\beta+\nu\sin 2\beta)}
\label{eq:sharp}
\end{equation}
The statistical significance of the signal of correlation in this
case is $$P=3.8\times 10^{-4}\%\ ,$$ taking into account the eight
parameters in the correlation ansatz. If the null hypothesis is assumed to be 
the von Mises distribution we find an even 
smaller value of $P$. The signal of correlation is clearly very strong
and cannot be dismissed as a statistical fluctuation. 
The parameters are given by
$k= -0.561\pm 0.071$, $\nu = -1.09\pm 0.17$, $c_1 = -1.71 \pm 0.32$, 
$c_2 = 9.64 \pm 0.47$, 
$c_3=2.89 \pm 0.55$, 
$|\bmath{\lambda}| =0.70\pm 0.15 $ and the axis parameters are given by 
\begin{equation}
RA= (0^{\rm h},\  3.7^{\rm m})\pm (0^{\rm h},\  46^{\rm m}),
\ DEC = -10.7^o\pm 12.2^o 
\label{eq:axis1}
\end{equation}
in the B1950 coordinate system.
Within errors this points opposite to the direction of the Virgo supercluster.
We see from the value of the 
parameter $c_1$ that the sources whose RM lies roughly within
the interval $c_2\pm c_3$ show a correlation with 
$\bmath{\lambda}\cdot {\bf R}$  which is opposite to
the remaining sources. The correlation term 
$\bmath{\lambda}\cdot {\bf R}(1+g(RM))$ has an opposite sign for these 
sources in comparison to the remaining sources. 
The axis parameters, Eq. \ref{eq:axis1}, obtained from this fit agree 
with the earlier fit, Eq. \ref{axis}, within errors.

The function $g(RM)$ essentially treats the sources whose RM lies in
the interval $c_2\pm c_3$ differently from the remaining sources. 
This interval is centered at the value $c_2$ which is very close to 
the position of the peak in the RM distribution as can be seen in 
Fig. \ref{hist_rm}. We may attribute this shift in the peak
to some bias in the extraction of RM or due to some systematic
physical effect such as the contribution due to the milky way. 
Hence if the difference $RM-c_2$ is small it may indicate that the
host contribution to RM is negligible. 
The introduction of the function $g(RM)$ in the correlation ansatz,
therefore, treats the sources with negligible contribution due to the
host galaxy differently than the remaining sources. 

The relationship between $\beta$ and the position vector \bmath{R} of the 
source indicated by Eq. \ref{joint_rm} may be expressed as follows,
\begin{equation}
{<(1+\nu\cos 2\beta)sin{\Omega}>\over <(1+\nu\cos 2\beta)cos{\Omega}>}
= { \bmath{\lambda}\cdot {\bf R}\over k} (1+g(RM)) \ 
\label{relationship}
\end{equation}
where the angular brackets indicate statistical averages. Since the
parameter $\nu \approx -1$, the function $(1+\nu\cos 2\beta)$ is
approximately 2 near the peak of the distribution of $\beta$ and zero near the 
tail. Hence this function weights the peak region more than the tail
in the statistical averages. For $\beta$ close to its peak value
$\pi/2$, we find that $\sin{\Omega}\approx -2(1-\nu)(\beta - \pi/2)$ and
$\cos{\Omega}\approx -1$. If $RM$ is not too close to its peak
value or more precisely $|RM-c_2| > c_3$ 
then $g(RM)\approx 1$ and hence we find the approximate relation
\begin{equation}
<\beta - \pi/2> \approx {1\over 2(1-\nu)}
{ \bmath{\lambda}\cdot {\bf R}\over k}\ ,
\label{relationship1}
\end{equation}
which is same as the relationship given in Eq. 
\ref{eq:anisotropy} with the \bmath{\Lambda} in Eq. \ref{eq:anisotropy}
identified with ${ \bmath{\lambda}\over (1-\nu)k}$ in Eq.
\ref{relationship1}.

\begin{figure}[h!]
\centering
\includegraphics[scale=1.00]{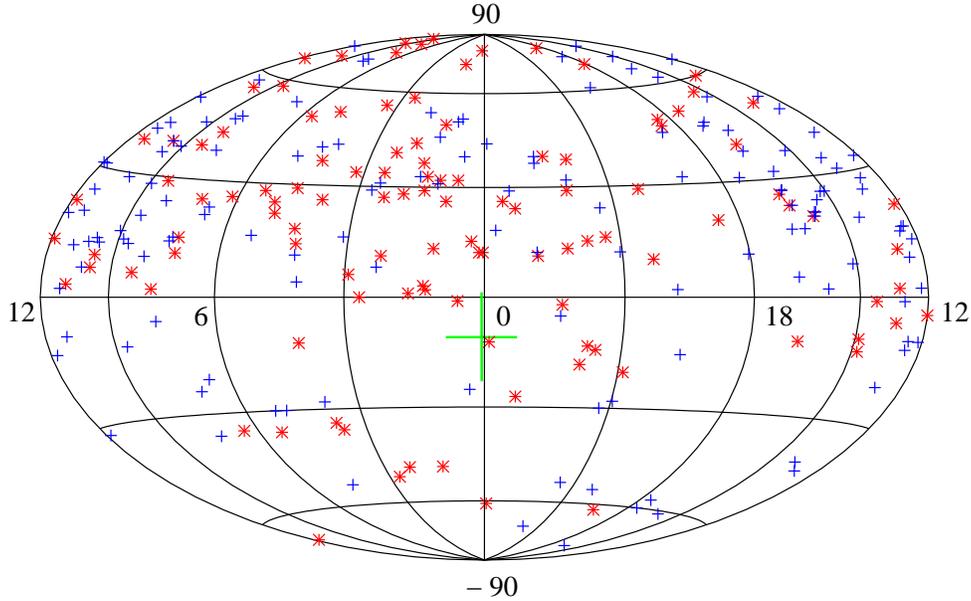}
\caption{Aithoff-Hammer scatter plot of the
correlation variable $(1+g(RM)) \sin\Omega $,
as defined in Eq. \ref{joint_rm}. Here red stars and blue plusses represent
sources for which $(1+g(RM)) \sin\Omega >0.1$ and 
$(1+g(RM)) \sin\Omega < -0.1$
respectively. The anisotropy axis, 
$[(0^{\rm h},\  3.70^{\rm m})\pm (0^{\rm h},\  46^{\rm m}),-10.7^o\pm 12.2^o]$,
along with its error bars is shown in green. 
}
\label{fig:AH1}
\end{figure}

\begin{figure}[h!]
\vskip -1.8in
\hskip -1.0in
\includegraphics[scale=0.85]{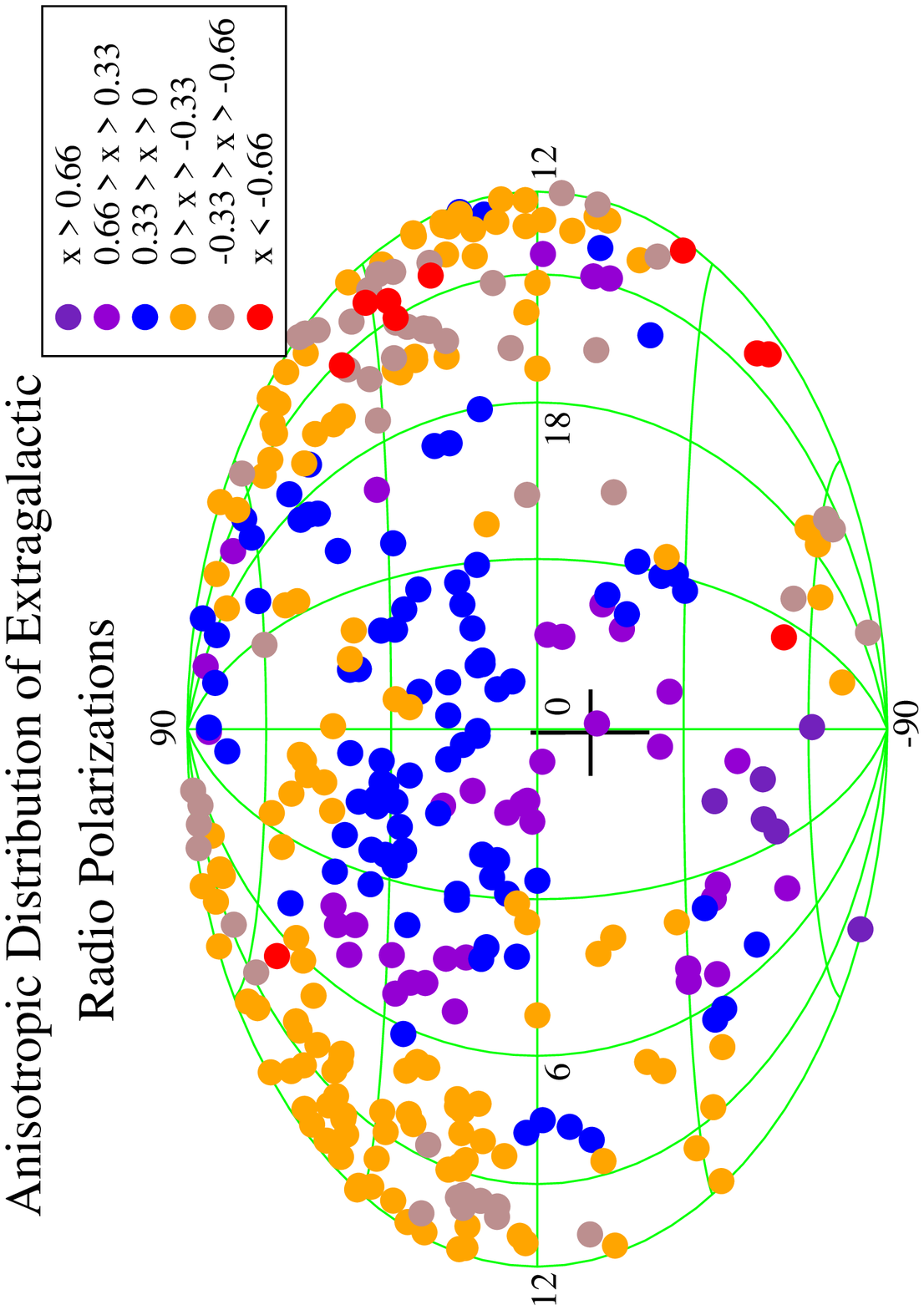}
\caption{Aithoff-Hammer scatter plot of the 
averaged correlation variable 
$x = \overline{(1+g(RM)) \sin\Omega} $,
as defined in Eq. \ref{joint_rm}. 
For each source located in the direction \bmath{R_1} 
we take the average over all the nearest neighbours which
lie within the cone defined by $\bmath{R_1}\cdot \bmath{R_2} < 0.95$, where
$\bmath{R_2}$ is the unit vector in the direction of any other source.
The anisotropy axis,
$[(0^{\rm h},\  3.70^{\rm m})\pm (0^{\rm h},\  46^{\rm m}),-10.7^o\pm 12.2^o]$,
is shown along with its error bars. 
Within errors it points opposite to the
Virgo supercluster. 
}
\label{fig:AH0_average}
\end{figure}

The anisotropic distribution of the offset angle obtained with the fit
Eq. \ref{joint_rm} may now be illustrated by making an Aithoff-Hammer
scatter plot of the correlation variable $(1+g(RM))\sin\Omega$. 
This is shown in fig. 
\ref{fig:AH1} where stars and plusses represent sources
for which $(1+g(RM)) \sin\Omega >0.1$ and $(1+g(RM)) \sin\Omega < 0.1$
respectively. We find that the stars  
are concentrated in the 
direction of the axis
with the plusses concentrated in the opposite
direction. The anisotropy in the angular distribution
of the offset angles is clear from this plot. The nature of the anisotropy
is also nicely illustrated by making a scatter plot after taking the local
averages. In fig. \ref{fig:AH0_average} we show the Aithoff-Hammer
scatter plot of the averaged correlation variable, 
$\overline{(1+g(RM)) \sin\Omega}$, at the location of each source. The average
is done over the nearest neighbours whose angular separation from the
source is less than $\cos^{-1} 0.95\approx 18^o$. The mean number of sources
which contribute to this average is found to be 12.6. 
The plot clearly shows
anisotropic nature of the distribution of the
offset angles. 

\bigskip
\noindent
{\Large\bf 3. Physical Origin of the Effect}

\medskip
From the results obtained above we see that
the offset angle $\beta$ shows some correlation with $RM$ but
the correlation is direction dependent. It is difficult to visualize
how such a direction dependent correlation can arise due to bias
in data. In all likelihood this is caused by some physical phenomenon. 
The offset angles of the different sources may be intrinsically 
correlated or the effect may arise due to propagation. The
possibility of intrinsic correlation can be further investigated 
by determining the correlation of the intrinsic position angle of polarization
$\chi$ and the orientation angle of the galaxy $\psi$ among the different 
sources. By using the statistical techniques 
used in (\huts 1998, \huts \& Lamy 2001, Jain, Narain \& Sarala   2003) 
to test of alignment of optical polarizations,
we find that $\chi$ and $\psi$ for different sources show no 
 alignment with 
one another. The $P$ value in this case is found to be larger
than 10 \% independent of the number of nearest neighbours used to 
test for alignment. 
Hence it is not possible to attribute the radio anisotropy to 
intrinsic alignment of radio polarization and it is likely to arise due to some 
propagation effect. Within conventional physics the corrections to the
Faraday polarization rotation effect are very small and hence it is 
unlikely that this anisotropy may be explained within this framework. 
The anisotropy may, therefore, be an indication of some new physical effect. 

One interesting possibility is
the presence of a very low mass background pseudoscalar field 
in the universe. We assume that at large redshifts, relevant for
the sources considered in the present data set, this field has a coherent
component which is approximately given as
\begin{equation}
\phi({\bf R}) \approx \phi_0\cos(\theta)(1+g[B_r({\bf R})])
\label{phi_X}
\end{equation}
where $\phi_0$ is a constant independent of redshift, 
$\theta$ is the angular position of the source measured with respect
to the prefered axis $\bmath\lambda$, $B_r$ is the radial component 
of the magnetic field in the coordinate frame with origin located at
the observation point i.e. earth and $g(B_r)$ is a gaussian defined
in Eq. \ref{eq:gaussian}. At the position of the earth we
assume that this field acquires a value $\phi_1$. The total rotation 
in the polarization angle due to this field is then equal to 
$\phi(\bf R) -\phi_1$ (Harari \& Sikivie, 1992). 
Since the rotation measure is proportional to 
$B_r$ we find that this effect can explain the observed correlation
given in Eq. \ref{joint_rm}. This mechanism requires a large 
scale anisotropy in the universe and provides the simplest explanation
of the observations. 

We next determine whether the observations can also be explained
within the framework of an isotropic universe. 
Although the correlation indicates the presence of a global anisotropy
it is possible that 
the correlation may arise due to the existence of a few patches of
length scales of order Gpc over which the pseudoscalar field 
shows a coherent dependence as a function of the angular coordinates.
For example there may exist two such patches, one in the direction of
the axis and another in the opposite direction. The observed anisotropy 
can arise if the pseudoscalar field in one of these patches is approximately
equal to the negative of its value in the other patch. In particular 
the observed correlation can be explained if 
$\phi({\bf R}) \approx \phi_0 (1+g(RM))$ in one of the patches and
$\phi({\bf R}) \approx -\phi_0 (1+g(RM))$ in the second patch.
We point out that although the existence of a large scale dipole distribution
of the background field violates the fundamental assumption of isotropy 
of the universe, the existence of the few patches of length scales of order
Gpc does not. Here we also assume that in eq. \ref{phi_X} $\phi({\bf R})$
depends on the magnitude $|{\bf B}|$ rather than on $B_r$. Such a 
functional dependence may also explain the observations provided we
attribute the RM contribution due to the host galaxy to the 
deviation of RM from its peak value, RM$_{\rm peak}$, in the RM distribution.  
As discussed earlier,
our fits also always select the parameter $c_2$ such
that it is approximately equal to RM$_{\rm peak}$. 
The shift in RM$_{\rm peak}$ from zero may be attributed to some local
effect such as the contribution due to the milky way or due to some bias in 
the extraction of RM.
The deviation of RM from its peak value,
RM-RM$_{\rm peak}$, gets contribution both from the host galaxy and the
milky way. Hence RM-RM$_{\rm peak}$
is directly proportional to the magnitude of the magnetic field in the
host galaxy, assuming that in a statistical sense all the components of
the background magnetic field have equal strength. 
This mechanism may explain the observed correlation
within the framework of an isotropic universe provided their exist 
patches of coherent background pseudoscalar field of length scales of
order Gpc. However coherence on such large scales is not expected 
in conventional
cosmology. Furthermore this explanation 
also requires an accidental correlation of two 
oppositely directed patches such that pseudoscalar field is
positive in one of these patches
and negative in the other. 

A large scale anisotropy has also been observed in the optical polarizations
from distant quasars (\huts 1998, \huts \& Lamy 2001, Jain et al. 2003). 
It was found  
that the optical polarizations are aligned with one another over 
very large distances of the order of Gpc. 
A very strong alignment was seen in a patch at large redshifts centered
at the Virgo supercluster (\huts 1998). 
Furthermore it was found
(Jain et al  2003) that the polarizations of the 
large redshift, $z\ge 1$, data sample are aligned over the entire 
celestial sphere. 
This effect is also not
easily explained within conventional cosmology/astrophysics.
The effect may have some relationship to the radio anisotropy discussed
in this paper since both the effects appear to be very strong in the 
direction of the Virgo supercluster.
As shown in Jain et al.  (2003) the optical alignment effect
can also be explained if we assume the existence of a light pseudoscalar.
The proposed explanation requires the existence of several patches of
large scale
magnetic fields of length scales of order Gpc. 
Furthermore the global alignment at 
large redshifts requires the existence of a magnetic field 
coherent over the entire universe.  
Alternatively this very large scale alignment can be explained
if the large redshift QSO's emit a significant
flux of light pseudoscalars at optical frequencies.

\bigskip
\noindent
{\Large\bf 4. Conclusions}

\medskip
To conclude, we find that there is considerable evidence for the 
presence of angular correlation in the radio offset angles which is
not easily explained within conventional cosmology/astrophysics. 
The effect may be explained by the presence of a hypothetical
pseudoscalar particle. The effect indicates the existence
of a global anisotropy in the universe.

\bigskip
\noindent{\bf Acknowledgements:} We thank Amir Hajian and John Ralston
for useful comments.

\bigskip
\noindent{\bf References}

\begin{itemize}
\item[]Bietenholz M. F.,   Kronberg P. P., 1984,  ApJ,
287, L1-L2 

\item[]
Birch P., 1982, Nature, 298, 451 

\item[] Broten N. W., MacLeod J. M., Vallee J. P., 1988,
Astrophysics and Space Science 141, 303

\item[] Dobado A.,  Maroto A. L.,
1997, astro-ph/9706044, Mod. Phys. Lett.  A 12,  3003

\item[]  Harari D.,  Sikivie P., 1992, {\it Phys. Lett.}  B 289, 67

\item[]
\huts D., 1998, {\it A \& A} { 332}, 410 

\item[]
\huts D.,  Lamy H., 2001, {\it A \& A}, { 367}, 381 

\item[]
Jain P., Ralston J. P., 1999, {\it Mod. Phys. Lett.}
{ A14}, 417 

\item[]
Jain P., Panda S., Sarala S., 2002, Phys. Rev. D { 66}, 085007; 
hep-ph/0206046

\item[]
Jain P., Narain  G., Sarala S., 2003, astro-ph/0301530, to be published
in MNRAS

\item[]
Jupp P. E.,  Mardia K. V., 1980, {\it Biometrika} { 67}, 163 

\item[]
Kendall D. G.,  Young A. G., 1984,  MNRAS, 207, 637 

\item[] Mansouri R., Nozari K., 1997,
gr-qc/9710028

\item[] Moffat J. W., 1997, astro-ph/9704300

\item[] Obukhov Y. N., 2000, astro-ph/0008106, Published in 
Colloquium on Cosmic Rotation: Proceedings edited by M. Scherfner, 
T. Chrobok and M. Shefaat 
(Wissenschaft und Technik Verlag: Berlin, 2000), 23-96.  

\item[] Obukhov Y. N., Korotky V. A., Hehl F. W., 1997, astro-ph/9705243

\item[]
Phinney E. S., Webster R. I., 1983, {\it Nature}
{ 301}, 735 

\item[]
Ralston J. P.,  Jain P., 1999, International
Journal of Modern Physics D 8, 537 

\item[]
Sarala S., Jain P., 2001, astro-ph/0007251, MNRAS { 328}, 623 

\item[]
Sarala S., Jain P., 2002, Journal of Astrophysics and Astronomy  23, 
137 

\item[] Surpi G. C., Harari D. D., 1999, astro-ph/9709087, 
ApJ { 515},  455

\item[]
Vall\'{e}e J. P., 1997, {\it Fundamentals of Cosmic Physics} { 19}, 1 

\item[]
Zeldovich Ya. B., Ruzmaikin  A. A., Sokoloff D. D., 1983,
{\it Magnetic fields in Astrophysics} (Gordon and Breach Science
Publishers, 1983).  
\end{itemize}
\end{document}